\documentclass[12pt]{article}
\usepackage{authblk}
\usepackage[bookmarksnumbered, colorlinks, plainpages]{hyperref}
\usepackage{amsmath, amsthm, amscd, amsfonts, amssymb, graphicx, color, booktabs}
\usepackage{subcaption}
\numberwithin{equation}{section}
\definecolor{email}{rgb}{0.00,0.00,0.84}
\begin{document}
\setcounter{page}{1}

\title{\large \bf 12th Workshop on the CKM Unitarity Triangle\\ Santiago de Compostela, 18-22 September 2023 \\ \vspace{0.3cm}
\LARGE Unitarity constraints \\and the dispersive matrix}

\author[1]{Guido Martinelli \thanks{guido.martinelli@roma1.infn.it}}
\affil[1]{Physics Department and INFN Sezione di Roma La Sapienza, Piazzale Aldo Moro 5, 00185 Roma, Italy}
\author[2]{Silvano Simula \thanks{silvano.simula@roma3.infn.it}}
\affil[2]{Istituto Nazionale di Fisica Nucleare, Sezione di Roma Tre, \newline Via della Vasca Navale 84, I-00146 Rome, Italy}
\author[3]{Ludovico Vittorio \thanks{ludovico.vittorio@lapth.cnrs.fr (speaker)}}
\affil[3]{LAPTh, Universit\'e Savoie Mont-Blanc and CNRS, \newline F-74941 Annecy, France}
\maketitle

\begin{abstract}
We present updated estimates of $\vert V_{cb} \vert$ and $R(D^{(*)})$ based on all the available theoretical and experimental data on semileptonic $B \to D^{(*)} \ell \nu_\ell$ decays. These values have been obtained by using the Dispersive Matrix method to describe the hadronic form factors. By analysing all the lattice data we get the theoretical values $R^{\rm th}(D) = 0.296 \pm 0.008$ and $R^{\rm th}(D^*) = 0.262 \pm 0.009$, which are consistent with the corresponding HFLAV averages at the $\simeq 2.0\,\sigma$ and the $\simeq 1.5\,\sigma$ level, respectively. Moreover, from a bin-per-bin study of the experimental data we obtain the values $\vert V_{cb} \vert = (41.0 \pm 1.2) \cdot10^{-3}$ from $B \to D$ decay and $\vert V_{cb} \vert = (39.92 \pm 0.64) \cdot10^{-3}$ from $B \to D^*$ one, whose differences with the latest inclusive determinations never exceed the $\simeq 2.5\,\sigma$ level.
\end{abstract} \maketitle

\section{Introduction}

\noindent In this proceeding we will present a new analysis of the Cabibbo-Kobayashi-Maskawa (CKM) matrix element $\vert V_{cb} \vert$ and of the Lepton Flavour Universality (LFU) ratios $R(D^{(*)})$. This updated study is, in fact, motivated by several novel experimental and theoretical results, respectively:
\begin{itemize}
\item[i)] the new measurements of the differential decay widths of semileptonic $B \to D^* \ell \nu_\ell$ decay, done by the Belle\,\cite{Belle:2023bwv} and Belle-II\,\cite{Belle-II:2023okj} Collaborations;
\item[ii)] the new results of the computation of the $B \to D^* \ell \nu_\ell$ form factors (FFs) on the lattice, performed by the FNAL/MILC Collaboration\,\cite{FermilabLattice:2021cdg}, by the HPQCD Collaboration\,\cite{Harrison:2023dzh} and by the JLQCD Collaboration\,\cite{Aoki:2023qpa}.
\end{itemize}
In what follows, we will adopt the Dispersion Matrix (DM) method of Refs.~\cite{DM1, DM2, DM3, DM4, DM5, DM6} to describe the hadronic FFs. We have explicitly verified that very similar results could, in principle, be obtained by using the more standard Boyd-Grinstein-Lebed (BGL) parametrization~\cite{Boyd:1997kz}.

\section{Basic ingredients of the DM method}

Let us briefly summarize here the main properties of the DM approach. The starting point is to associate to a generic FF, call it $f$, the matrix
\begin{equation}
\label{eq:Delta2}
\mathbf{M} = \left( 
\begin{tabular}{ccccc}
   $\chi$ & $\phi f$ & $\phi_1 f_1$ & $...$ & $\phi_N f_N$ \\[2mm] 
   $\phi f$ & $\frac{1}{1 - z^2}$ & $\frac{1}{1 - z z_1}$ & $...$ & $\frac{1}{1 - z z_N}$ \\[2mm]
   $\phi_1 f_1$ & $\frac{1}{1 - z_1 z}$  & $\frac{1}{1 - z_1^2}$ & $...$ & $\frac{1}{1 - z_1 z_N}$ \\[2mm]
   $... $  & $...$ & $...$ & $...$ & $...$ \\[2mm]
   $\phi_N f_N$ & $\frac{1}{1 - z_N z}$ & $\frac{1}{1 - z_N z_1}$ & $...$ & $\frac{1}{1 - z_N^2}$
\end{tabular}
\right),
\end{equation}
where we have introduced the conformal variable $z$ defined as 
\begin{equation}
\label{conf}
z(t) = \frac{\sqrt{t_+ -t} - \sqrt{t_+ - t_-}}{\sqrt{t_+ -t} + \sqrt{t_+ - t_-}},\,\,\,\,\,\,\,\,\,\,\,\,\,\,\,\,\,\,\,\,\,t_{\pm}=(m_B \pm m_{D^{(*)}})^2.
\end{equation}
In the previous expression, $\phi_i f_i \equiv \phi(z_i) f(z_i)$ (with $i = 1, 2, ... N$) are the known values of the quantity $\phi(z) f(z)$ corresponding to the values $z_i$ at which the FFs have been computed on the lattice (the explicit expressions of $\phi(z)$ for each FF can be found in Ref. \cite{DM3}). Finally, the susceptibility $\chi(q_0^2)$ is related to the derivative with respect to $q_0^2$ of  the Fourier transform of suitable Green functions of bilinear quark operators and follows from the dispersion relation associated to a particular spin-parity quantum channel. Their non-perturbative values can be found in \cite{DM2} for $b \to c$ quark transitions for $q_0^2=0$. Note that a new computation of these quantities on the lattice is on-going, aiming at determining them with higher precision \cite{Melis:2024wpb}.

At this point, the determinant of the matrix $\mathbf{M}$ has to be, by construction, semi-positive definite, $i.e.$ $\det \mathbf{M} \geq 0$. The positivity of the determinant, which acts as a \emph{unitarity filter}, allows then to compute the lower and the upper bounds of the FF of interest for each generic value of $z$, $i.e.$
\begin{equation}
f_{\rm lo}(z) \leq f(z) \leq f_{\rm up}(z).
\end{equation}
To be more quantitative, we have that \cite{DM1}
\begin{equation}
   \beta - \sqrt{\gamma} \leq  f \leq \beta + \sqrt{\gamma} ~ , ~
    \label{eq:bounds}
\end{equation} 
where (after some algebraic manipulations)
\begin{equation*}
      \beta = \frac{1}{\phi(z)d(z)} \sum_{j = 1}^N f_j \phi_j d_j \frac{1 - z_j^2}{z_0 - z_j},\,\,\,\,\,\,\,\,\,\, \gamma  =   \frac{1}{1 - z_0^2} \frac{1}{\phi(z)^2 d(z)^2} \left( \chi - \overline{\chi} \right),   
\end{equation*} 
\begin{equation*}
       \overline{\chi}  =  \sum_{i, j = 1}^N f_i f_j \phi_i d_i \phi_j d_j \frac{(1 - z_i^2) (1 - z_j^2)}{1 - z_i z_j}.
\end{equation*}
Here $d(z),\,d_i$ are kinematical functions. Unitarity is satisfied only when $\gamma \geq 0$, which implies $\chi \geq \overline{\chi}$. While the values of $\beta$ and $\gamma$ depend on $z$, the value of $\overline{\chi}$ does not depend on $z$. In other words, $\overline{\chi}$ depends only on the set of input data and, thus, the unitarity condition $\chi \geq \overline{\chi}$ does not depend on $z$.

In the following Section, we will describe in detail the applications of the DM method to semileptonic $B \to D^{(*)} \ell \nu$ decays. We will show updated values of $\vert V_{cb} \vert$ and $R(D^{(*)})$, obtained by considering all the available data both from the lattice and from the experiments. Although in this proceeding we will focus on semileptonic $B \to D^{(*)}$ decays within the Standard Model (SM), it is worth mentioning that, from the methodological point of view, the DM results for the hadronic FFs can be used also for investigations beyond the SM. To be more specific, they can be directly incorporated in global New Physics (NP) fits, as recently done in \cite{NPBDstar1, NPbs}. 

\section{Updated values of $\vert V_{cb} \vert$ and $R(D^{(*)})$}

Let us begin with the analysis of semileptonic $B \to D$ decay, which has been studied in detail in Ref.~\cite{DM3} in the context of the DM method. Starting from the results of the computation of the relevant FFs on the lattice by FNAL/MILC Collaborations \cite{MILC:2015uhg} at high momentum transfer, we use the DM approach to describe the FFs in a non-perturbative and model-independent way in the entire kinematical region. In this way, we can firstly compute the fully-theoretical estimate $R(D) = 0.296 \pm 0.008$. Then, we determine bin-per-bin estimates of $\vert V_{cb} \vert$ through the experimental determinations of the differential decay width in Ref.~\cite{Belle:2015pkj}. A correlated average of the bin-per-bin values gives the final result $\vert V_{cb} \vert = (41.0 \pm 1.2) \cdot10^{-3}$. 

\begin{figure}
    \centering
    \begin{subfigure}{.56\textwidth}
        \centering
        \includegraphics[width=.9\linewidth]{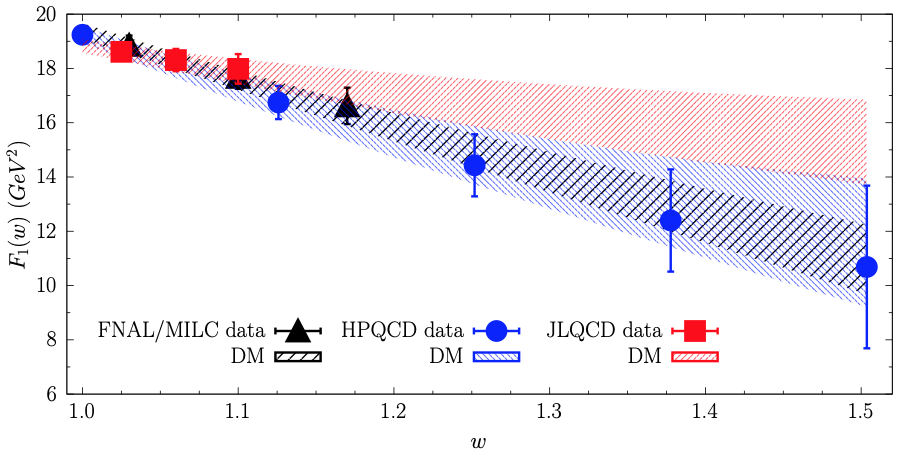}
        \caption{}
    \end{subfigure}%
    \begin{subfigure}{.46\textwidth}
        \centering
        \includegraphics[width=.9\linewidth]{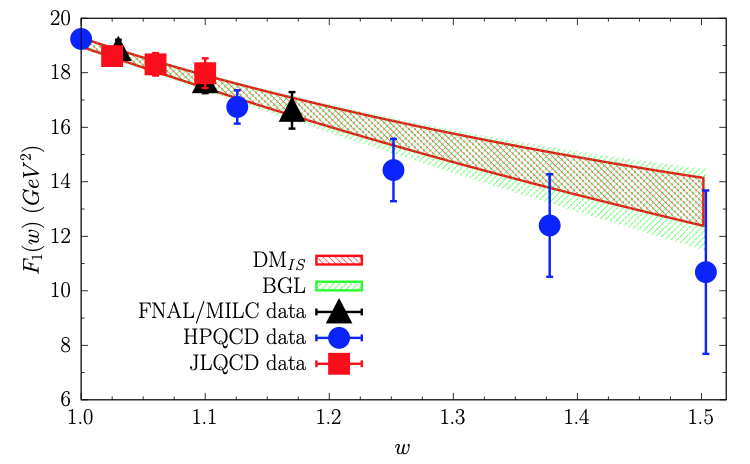}
        \caption{}
    \end{subfigure}
    \caption{The FF $F_1(w)$ entering in the semileptonic $B \to D^* \ell \nu_\ell$ decays, computed by studying separately the FNAL/MILC (triangles), HPQCD (circles) and JLQCD (squares) datasets (left panel) or by developing directly a joint analysis through an IS procedure (right panel).}
    \label{fig:Fig1}
\end{figure}

For what concerns instead semileptonic $B \to D^*$ decay, we can study each of the lattice datasets mentioned in the Introduction separately or we can use them in a joint analysis, as done in Ref.~\cite{DM7}. In Figure \ref{fig:Fig1} the reader can see the results of the application of these two strategies to one of the relevant FFs, namely $F_1(w)$. Let us highlight here that, since the unitary filters become more and more selective by increasing the number of input data, it is mandatory to use an Importance Sampling (IS) procedure to complete the joint analysis, $i.e.$ by following the procedure explained in Ref.~\cite{pion_IS}. 

These two sets of bands of the FFs  can be used to compute fully-theoretical values of the LFU ratio $R(D^*)$ and of the polarization observables. In the former case, our final value reads $R(D^*) = 0.262 (9)$, obtained through an average of the three $R(D^*)$ estimates coming from the analysis of each lattice dataset separately (a PDG scale factor of 1.8 has been taken into account). Had we used the $DM_{IS}$ bands to do this computation, we would have obtained $R(D^*) = 0.259 (5)$, in perfect agreement with the previous value. This exercise can be repeated for every physical quantity of interest for phenomenology. To be more specific, we have done it for  the $\tau$-polarization $P_{\tau}(D^*)$, for the longitudinal $D^*$-polarization fraction with heavy and light charged leptons, $F_{L, \tau}$ and $F_{L, \ell}$ ($\ell = e, \mu$) respectively, and for the forward-backward asymmetry $A_{FB, \ell}$. To summarize our findings, no significative difference can be found among theory and measurements for any of these quantities with the only exception of $F_{L, \ell}$. This discrepancy can thus point towards the existence of possible NP effects coupled to the light generations of leptons, as already investigated in Ref.~\cite{NPBDstar1}.

\begin{figure}
    \centering
    \begin{subfigure}{.56\textwidth}
        \centering
        \includegraphics[width=.8\linewidth]{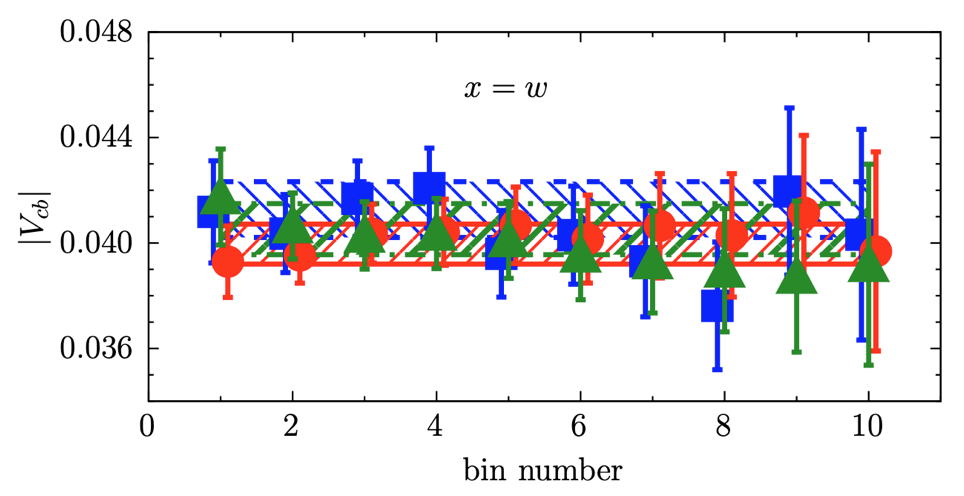}
        \caption{}
    \end{subfigure}%
    \begin{subfigure}{.46\textwidth}
        \centering
        \includegraphics[width=0.98\linewidth]{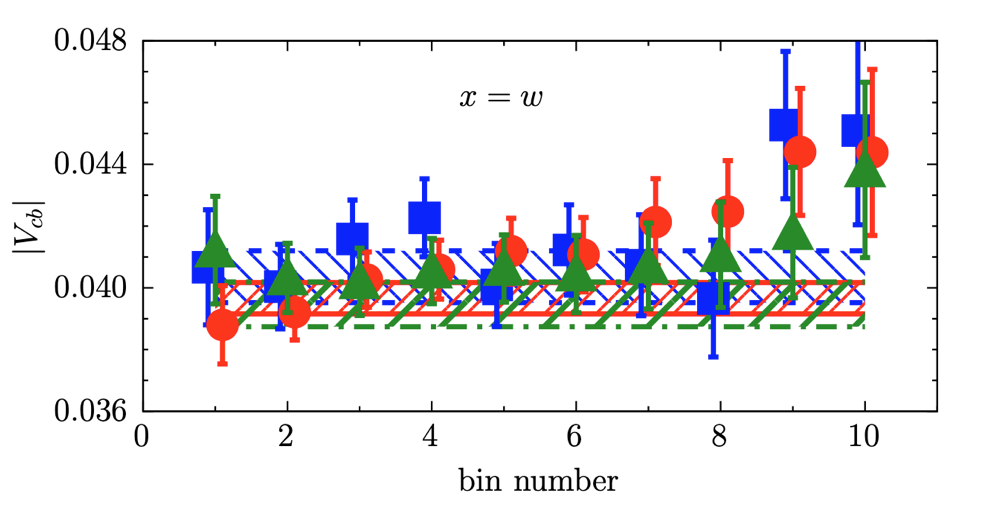}
        \caption{}
    \end{subfigure}
    \caption{Bin-per-bin estimates of $\vert V_{cb} \vert$ in the recoil distribution by using the experimental data from Refs.\,\cite{Belle:2023bwv} (blue squares),\,\cite{Belle-II:2023okj} (green triangles) and \cite{Belle:2018ezy} (red circles). The FFs have been determined by using as input the JLQCD lattice data (left panel) or by adopting the FFs as extrapolated through the $DM_{IS}$ method. The horizontal blue, green and red bands correspond to the correlated averages of these values.}
    \label{fig:Fig2}
\end{figure}

For what concerns $\vert V_{cb} \vert$, our proposal is to consider each measurement as a determination of $\vert V_{cb} \vert$ \emph{per se}, as done for semileptonic $B \to D$ decay. As an explicit example, in Figure \ref{fig:Fig2} we show the $\vert V_{cb} \vert$ distribution in the recoil variable by using as input the JLQCD lattice data (left panel) or by adopting the FFs as extrapolated through the $DM_{IS}$ method. While in the former case the distribution is completely flat, in the latter one $\vert V_{cb} \vert$ seems to increase at high recoil. Since this behaviour is not possible in the SM, this is a potential signal that either some systematic effects are still present in theoretical/experimental data or some NP contributions are at work \cite{NPBDstar1, NPBDstar2, NPBDstar3, NPBDstar4}. By combining together all the $\vert V_{cb} \vert$ averages (corresponding to each lattice dataset and to each experimental dataset) we finally obtain  $\vert V_{cb} \vert = (39.92 \pm 0.64)\cdot 10^{-3}$, which is compatible respectively at the $\simeq 2.5\,\sigma$ and $\simeq 2.0\,\sigma$ level with the most recent inclusive determinations $\vert V_{cb} \vert^{\rm{incl}} = (41.97 \pm 0.48) \cdot 10^{-3}$~\cite{Finauri:2023kte} and $\vert V_{cb} \vert^{\rm{incl}} = (41.69 \pm 0.63) \cdot 10^{-3}$~\cite{Bernlochner:2022ucr}. Had we used directly the theoretical FFs obtained through the $DM_{IS}$ method, we would have found $\vert V_{cb} \vert \cdot 10^{3} = 39.87 \pm 0.55$, which is perfectly compatible with our reference value. Let us finally highlight that our result is in very nice agreement with the exclusive estimate $\vert V_{cb} \vert \cdot 10^{3} = 40.3 \pm 0.5$ recently obtained in Ref. \cite{Ray:2023xjn}.

\section{Conclusions}

In this proceeding we have reviewed the results of the application of the DM method to semileptonic $B \to D^{(*)} \ell \nu_\ell$ decays, obtained by using the most recent data from lattice and from experiments. Our theoretical estimates of the LFU ratios $R(D^{(*)})$ are in agreement with the corresponding HFLAV averages within 2.0$\sigma$, while an interesting discrepancy among theory and measurements afflicts $F_{L, \ell}$. For what concerns our determinations of $\vert V_{cb} \vert$, their differences with the inclusive values never exceed the $\simeq 2.5\,\sigma$ level. A combination of our numbers with the $\vert V_{cb} \vert$ values determined from $B_s \to D_s^{(*)} \ell \nu_\ell$ decays will be used in the near future for updated analyses of the Unitarity Triangle \cite{UTfit}.


\bibliographystyle{amsplain}

\end{document}